\documentclass[preprint,12pt]{elsarticle}
\usepackage{amssymb}
\usepackage{amsmath}
\usepackage{newunicodechar}
\usepackage[colorlinks,            
linkcolor=blue,            
anchorcolor=blue,            
citecolor=blue,
urlcolor=black]{hyperref}
\usepackage[normalem]{ulem}
\usepackage{multirow}
\usepackage{booktabs}
\usepackage{tabularx}
\usepackage{makecell}
\usepackage[dvipsnames]{xcolor}

\newcounter{bla}

\journal{Computer Physics Communications}

\begin{document}
\begin{frontmatter}

\title{Hybrid Multi-Head Physics-informed Neural Network for Depth Estimation in Terahertz Imaging}

\author[1,2,3]{Mingjun Xiang}
\author[2]{Hui Yuan}
\author[4,1]{Kai Zhou}
\author[2]{and Hartmut G. Roskos}

\address[1]{Frankfurt Institute for Advanced Studies (FIAS), 60438 Frankfurt am Main, Germany}
\address[2]{Physikalisches Institut, Goethe-Universität Frankfurt am Main, 60438 Frankfurt am Main, Germany}
\address[3]{Xidian-FIAS International Joint Research Center, 60438 Frankfurt am Main, Germany}
\address[4]{School of Science and Engineering, The Chinese University of Hong Kong, Shenzhen, P.R. China}

\begin{abstract}
Terahertz (THz) imaging is a topic in the field of optics, that is intensively investigated not least due to its potential for recording three-dimensional (3D) images, useful e.g., for the detection of hidden objects, nondestructive testing, and radar-like imaging in conjunction with automotive systems. 
Depth information retrieval is a key factor to %restore 
recover the three-dimensional shape %appearance 
of objects. Impressive results for depth determination %extraction 
in the visible and infrared spectral range have been demonstrated through deep learning (DL). Among them, most DL methods are merely data-driven, lacking relevant physical priors, %which leads to the need for a large amount of experimental data to optimize their weights and biases. 
which thus requires %request for 
a large amount of experimental data to train the DL models.
However, acquiring large training data in the THz domain is challenging
%obtaining a sufficient number of training data is challenging in the THz domain 
due to the time-consuming data acquisition process and environmental and system stability requirements during this lengthy process.
To overcome this limitation, this paper incorporates a complete physical model representing the THz image formation process into a DL network (NN) to retrieve the depth information of objects. The most significant advantage is using the DL NNs without pre-training, eliminating the need for tens of thousands of labeled data. Through experimental validation, we demonstrate that by providing diffraction patterns of planar objects, with their upper and lower halves sequentially masked to overcome the trapping of the NN's computational iterations in local minima, the proposed physics-informed NN can automatically reconstruct the depth of the object through interaction between the NN and the physical model.  The obtained results also represent the initial steps towards achieving fast holographic THz imaging using reference-free beams and low-cost power detection.
\end{abstract}
\end{frontmatter}

\section{Introduction}
% Importance of THz
THz radiation refers to electromagnetic radiation in the frequency range from %approximately 
0.1 THz to 10 THz, corresponding to wavelengths of 3 mm to 30 $\mu$m. THz imaging technology \cite{valusis_roadmap_2021}, as an important %promising 
direction in THz research, primarily benefits from the unique physical properties of this frequency band: firstly, it possesses strong penetration capability, enabling %easy 
transmission through non-metallic and non-polar materials such as ceramics, plastics, and foam, which are commonly opaque to visible and infrared light; secondly, THz waves exhibit low photon energy, thus avoiding harmful ionization effects; thirdly, many molecules display distinct absorption and dispersion characteristics in the THz frequency range, allowing the establishment of molecular fingerprint spectra for substance identification  \cite{zhang2010introduction, redo2008terahertz, PAWAR2013157}. 
Moreover, THz waves demonstrate high sensitivity to water, making them particularly suitable for analyzing material hydration levels \cite{Banerjee08}. Over the past two decades, advancements in electronic and photonic technologies have propelled the rapid development of THz imaging technology \cite{Friederich2011, Petrov2020, Valzania2018}, expanding its application areas into fields such as non-destructive testing \cite{amenabar2013introductory, Hasegawa2003}, quality monitoring \cite{ellrich2020terahertz}, security screening \cite{Federici_2005, Friederich2011}, biomedical imaging \cite{yang2016biomedical}, and sensing for robotics and vehicle control \cite{jasteh2015low}, where it has achieved notable successes, demonstrating advantages unmatched by traditional imaging techniques.

% Conventional methods for depth retrieval
Depth information plays a crucial role in THz imaging; however, due to the %lack of suitable hardware, 
large wavelength, the construction of THz multi-pixel detector arrays poses challenges, making it %difficult 
impossible to achieve a camera with a detector array consisting of millions of pixels akin to cameras in the visible domain \cite{rogalski2011terahertz, Hadi2012, Zdane2012}. Therefore, although there are well-established algorithms for depth restoration in the visible light domain \cite{8627998}, applying these algorithms to the THz domain remains quite challenging. 
At the current technological level, determining %estimating 
depth in THz imaging primarily relies on one of the following methods: focus scan measurements \cite{Perraud:19}, time-of-flight detection \cite{kim2022terahertz}, holographic detection and tomography. Focus measurement techniques determine the distance of target objects by adjusting the beam focus, but this often requires multiple scans of the target, thus consuming considerable time and resources. Time-of-flight utilizes the propagation time of THz waves to estimate the distance to the object's surface, but precision may be compromised in complex scenes or heavily occluded scenarios. A successful, industry-deployed frequency-domain variant of time-of-flight imaging is frequency-modulated continuous wave (FMCW) imaging \cite{Cristofani2014, Bac2017}, a powerful technique which relies on heterodyne detection \cite{Glaab2010}. The latter involves mixing the transmitted signal from the object with a local reference signal to achieve phase information, but this requires a fairly complex and expensive measurement system. Holographic detection schemes such as digital holography \cite{Locatelli2015, Petrov2016, Siemion2021} and Fourier imaging \cite{Yuan19, yuan2023a} obtain the distance information from the spatial distribution of the phase information across the detection area. The detection schemes always require the use of a reference beam, which usually is a substantial burden on the total power budget. Tomography is a widely used depth-resolving measurement technique \cite{Guillet2014}, but, due to the lack of THz cameras, it is very time-consuming.
% and is susceptible to environmental noise and stray signals. 
Therefore, exploring more efficient methods for depth information acquisition is imperative for advancements in THz imaging technology. 

% DL methods
Recently, DL techniques have emerged as promising and effective methods in the field of solving inverse problems  \cite{Pang:2016vdc,Shi:2021qri,Wang:2021jou,Shi:2022yqw,Soma:2022vbb}, primarily relying on supervised DL approaches that utilize labeled experimental data \cite{boominathan2018phase,deng2020learning,ju2018feature}. Nevertheless, the time-expensive acquisition of THz images poses a challenge for supervised training with experimental data.

% Our method
In this paper, we propose a novel hybrid multi-head physics-informed NN (PINN) for depth estimation in THz holographic imaging. This method incorporates the multi-input convolutional neural network (CNN) \cite{ronneberger2015u} and fully-connected NN with a physical prior knowledge, specifically, the angular spectrum theory \cite{goodman2005introduction, Xiang2022}. Hence, there is no need for thousands of labeled data to train the network. Instead, the model is trained with only two diffraction patterns, each with either the upper or lower portion masked, serving as inputs. The NN weights and bias factors are then optimized through the interaction between the NN and the physical model, ultimately yielding a feasible model solution that satisfies the imposed physical constraints.

\section{Methods}

\begin{figure*}[htb]
\centering
\includegraphics[width=\linewidth]{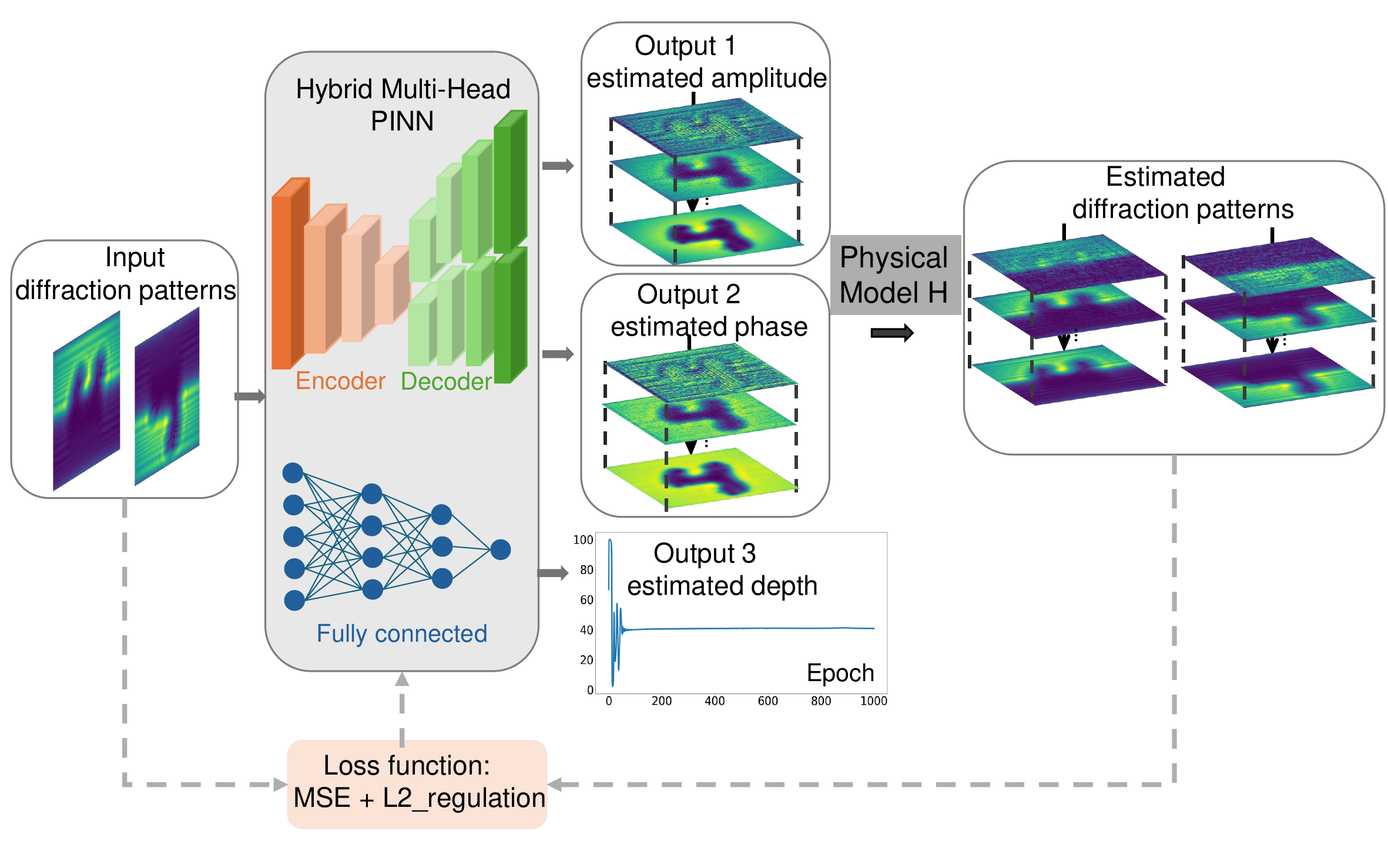}
\caption{Schematic pipeline illustration of the proposed depth estimation algorithm. The input of the CNN are two measured field-amplitude diffraction patterns, each with half of the object masked complementarily. The output consists of the estimated (1) the phase and, (2) amplitude maps, along with (3) the depth of the object. Subsequently, these outputs are numerically propagated through the physical model $H$ to simulate the diffraction and measurement processes, estimating the corresponding diffraction patterns. The MSE between the input patterns and estimated patterns, along with L2 regularization, serve as the loss function for optimizing the network parameters. }
\label{fig:false-color1}
\end{figure*}

Figure~\ref{fig:false-color1} displays the proposed hybrid multi-head PINN algorithm, demonstrated using handwritten digits as the object for example. The objects are planar, and placed in the experiments into the collimated THz beam perpendicular to the optical axis. 
Note that no pre-training on large labeled training datasets is required. The network relies solely on two amplitude diffraction patterns as input, each obtained by masking the upper or lower halves of the to-be-measured object and propagating the transmitted waves over a distance $d$. Experimentally, the amplitude diffraction pattern is obtained either directly by heterodyne measurements or as the square root of intensity maps recorded by power detection. The first head (output) of the PINN is estimating the amplitude map, while the second is for the estimated phase of the transmission function of the imaged objects. The two output paths share the same downsampling CNN layers considering that the same object information is analyzed, which is inspired by the U-Net structure \cite{alzubaidi2021review}. Within the upsampling layers, these two paths separate from each other. The third output of the PINN predicts the value of the depth, estimated with a fully connected network structure. CNNs are widely used in image reconstruction tasks because of their ability to efficiently capture local features through shared weights and local receptive fields, their robustness to translation invariance\cite{Zhou:2018ill}, and their hierarchical feature extraction capability\cite{alzubaidi2021review}. In our scenario, the amplitude within the masked metal area is set to zero, while the phase, instead of being made random, is also set to zero for simplicity. %For simplification of the computation, we set the phase within the masked metal area to be 0 as well.
Subsequently, the physical model $H$ is applied to convert network outputs into the corresponding diffraction patterns. The mean squared error (MSE) between inputs (the actual diffraction patterns) and the estimated diffraction patterns, coupled with L2 regularization \cite{cortes2012l2} (to help avoid overfitting), guides the optimization of the weight and bias parameters of the network via gradient descent. In image processing, MSE is commonly used to measure the quality of the reconstructed image compared to the ground truth. It represents the Euclidean distance between images and is simple and efficient to evaluate, making it suitable for large-scale image processing tasks. Additionally, MSE is a convex function (in terms of the network output) and possesses good mathematical properties \cite{4775883}. The L2 regularization aims to constrain the model's complexity by incorporating weight decay, thereby enhancing the model's generalization ability and mitigating overfitting which may occur if the object itself is of low complexity \cite{cortes2012l2}. This loss function pushes the calculated diffraction patterns to approach the input patterns iteratively with training, also achieving simultaneous searches for amplitude, phase, and depth information to converge toward consistent feasible solutions. Notably, since all parameters enter into the physical model $H$, %are considered, 
the proposed algorithm is universally applicable to holographic systems, irrespective of whether they involve phase objects or not.

As mentioned, the physical model $H$ simulates the THz imaging process in the experiment. If a planar object is illuminated by a collimated continuous-wave beam with a given cross-sectional amplitude profile and a zero phase offset, the complex-valued field amplitude right behind the object can be written as
\begin{equation}
E(x,y,z=0) = A(x,y,0) e^{i\phi(x,y,0)} ,
\label{eq:refname1}
\end{equation}
where $A(x,y,0)$ and $\phi(x,y,0)$ are the amplitude and the phase of the transmitted beam. Over a distance $d$, diffraction reshapes the field to \cite{goodman2005introduction}
\begin{equation}
E(x,y,z=d) = \iint \hat{E}_0(f_x,f_y)\,G\,e^{i2\pi(f_x x+f_y y)} df_x\,df_y
\label{eq:refname2}
\end{equation}
with $G=e^{ikd\sqrt{1-\lambda^2f_x^2-\lambda^2f_y^2}}$ the wave propagation function, and $\lambda$ the wavelength, $\hat{E}_0$ the spatial Fourier transform of $E(x,y,z=0)$ with $f_x=x/(\lambda d)$ and $f_y=y/(\lambda d)$ as the spatial frequencies in the $x$ and $y$ directions. The diffraction pattern is the absolute value of the propagated field %\HY{(here you use the amplitude as the diffraction pattern, well for real direct power detection it is the power which is the square of the amplitude. For all your experiments you do heterodyne detection and only take the amplitude right?) yes I only use amplitude from heterodyne detection}
\begin{equation}
A(x,y,z=d) = \left|E(x,y,z=d)\right| = H(\phi(x,y,0), A(x,y,0), d)),
\label{eq:refname3}
\end{equation}
where $H$ represents the mapping function relating the object's transmission function and the wave propagation to the diffraction pattern $A(x,y,z=d)$. 

For a conventional fitting using supervised NNs other than PINN, one usually attempts to derive a mapping function $R_\theta$ ($\theta$ denotes the network weights and bias parameters with $\theta^*$ is their optimum) from a large number of labeled data $(A(x,y,z=d)_k, d_k)$ contained in a labeled training set $S_T = {(A(x,y,z=d)_k, d_k), k=1,2,..., K}$, with its optimum defined as:
\begin{equation}
\theta^* = \mathop{\arg\min}_{\theta} \|R_\theta(A(x,y,z=d)_k)-d_k\|^2 \qquad\forall(d_k,A_k)\in S_T.
\label{eq:refname4}
\end{equation}

Instead, in the PINN proposed here, the loss function is formulated by combining the physical model $H$ and the network output $R_{\theta}$ as
\begin{equation}
loss = \|H(R_\theta(A(x,y,z=d)))-A(x,y,z=d) \|^2 + \lambda \sum_{i=1}^m \theta^2 \,,
\label{eq:refname5}
\end{equation}
where the diffraction pattern $A(x,y,z=d)$ is the input of the NN. The regularization term $\lambda\sum_{i=1}^m\theta^2$
is defined as the Euclidean norm or L2 norm of the weight matrix, which is the sum of all squared weight values. Here, we set the value of regularization strength $\lambda$ as 0.01, a small change of which does not induce different results in this study. 
In this way, the PINN is optimized with gradient descent (Adam optimizer is used) with the above objective
\begin{equation}
\theta^* = \mathop{\arg\min}_{\theta}(loss) \,.
\label{eq:refname6}
\end{equation}
When the optimization is complete, the resulting mapping function can then be used to reconstruct the depth with its third output:
\begin{equation}
 \tilde d= R_{\theta^*}(A(x,y,z=d)) \,.
\label{eq:refname7}
\end{equation}
%\HR{[Is this equation complete/correct? Isn't there a projection onto $d$ missing? $R_{\theta}$ is the entire weight and bias parameter set, and not identical to a distance.]}\KZ{$R_{\theta}$ is the mapping function with the 3rd output to be depth, and $\theta$ is the entire weight and bias params}

\begin{figure*}[htb]
\centering
\includegraphics[width=\linewidth]{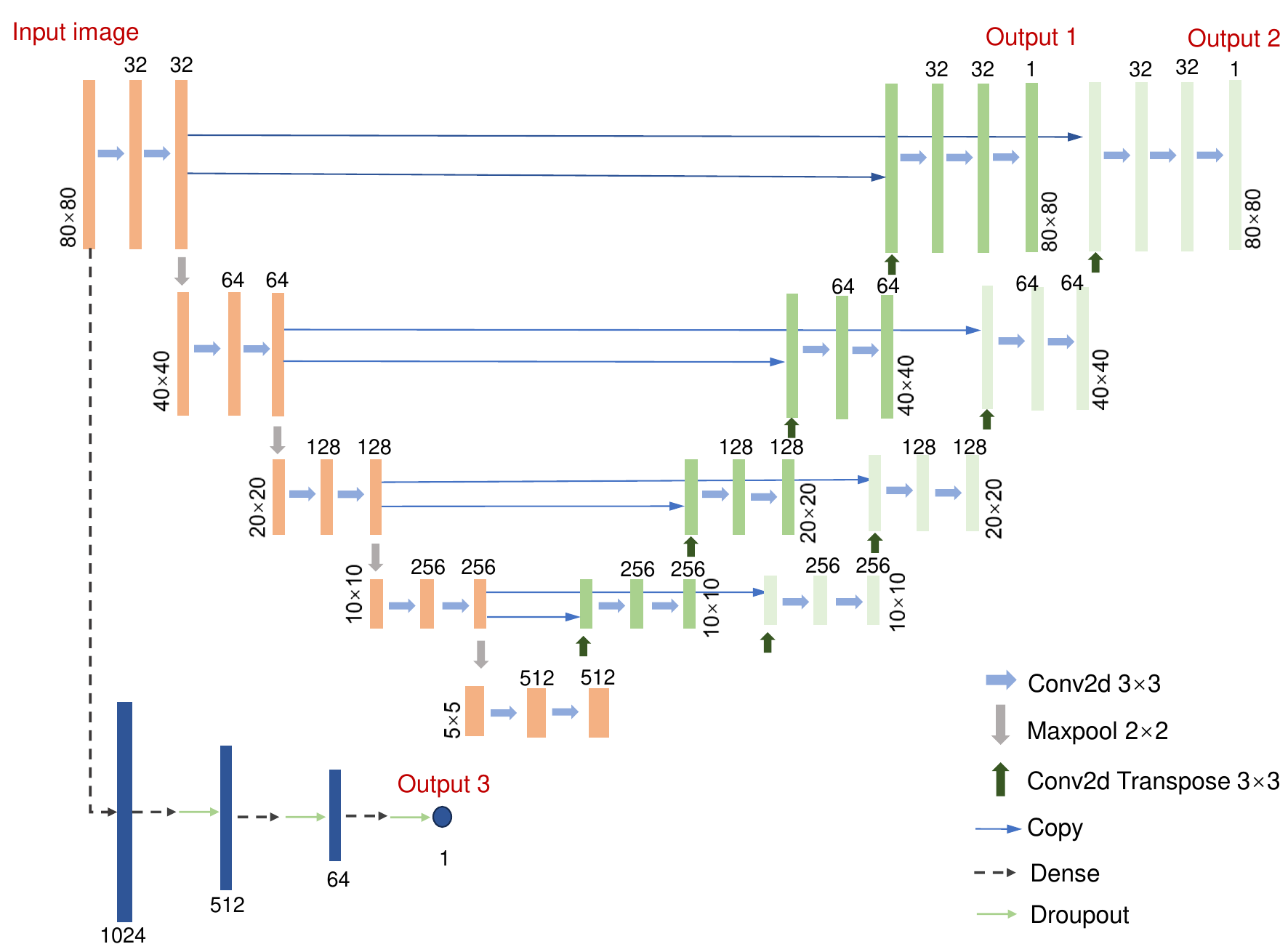}
\caption{Structure of the proposed hybrid multi-head PINN}
\label{fig:false-color7}
\end{figure*}

In designing the network, we opted for shared downsampling for output 1 (amplitude map) and output 2 (phase map), which start to separate from the upsampling layers, as shown in Fig.~\ref{fig:false-color7}. This design offers clear advantages over two common alternatives: Separating at the final downsampling layer or keeping the phases entirely separate.
We experimentally tested all three approaches.
Table~\ref{tab:comp} compares the performances of the three approaches. Using six real datasets, our design achieved an average MSE of 0.0024, outperforming the other two designs with MSEs of 0.0731 and 0.0065. Our network converged in 10 minutes at 1000 epochs, while the total-separation method took 12 minutes, and the final-layer-separation method took 20 minutes and needed 2000 epochs.
\begin{table*}[htb]
\centering
\caption{\bf Comparison of different network designs}
\begin{tabular}{ccc}
\toprule 
Method &  \makecell{Quality \\ (MSE)}  & \makecell {Convergence \\ time} \\
%Method  & (MSE) & time \\
\midrule
Fully Separate Downsampling and Upsampling & 0.0731 & $\sim$12~min \\
Separate Only at Final Downsampling Layer & 0.0065 & $\sim$20~min \\
Shared Downsampling, Separate Upsampling & 0.0024  & $\sim$10~min \\
\bottomrule
\end{tabular}
  \label{tab:comp}
\end{table*}

The results can be summarized as follows.  Firstly, compared to separating at the final downsampling layer, our approach significantly accelerates convergence. Sharing the downsampling segment allows the outputs to share information early, facilitating stable and rapid feature learning and %This results in faster, more stable training, 
avoiding the slow and inaccurate learning seen when separation occurs only at the final layer. Shared downsampling uses common feature extraction layers efficiently and reduces redundant computations. % and shortening runtime. 
Information sharing also enhances feature consistency and complementarity, improving overall accuracy.
Secondly, compared to entirely separate downsampling and upsampling phases, our design reduces demands on resources, computational overhead and training time. %Independent structures for each output require more resources and longer training times. 
%\HR{\sout{This design ensures effective feature extraction, accuracy, and computational efficiency, achieving stable and precise results quickly.} [This is repetitive.}]

The architecture of the proposed hybrid multi-head PINN algorithm was implemented using TensorFlow, an open-source DL software package \cite{abadi2016tensorflow}. We adopted the Adam optimizer \cite{kingma2014adam} with a learning rate of 0.0001 to optimize the network weights and biases. We added uniformly distributed noise between 0 and $\frac{1}{30}$ to the fixed input diffraction pattern in every optimization step to achieve better convergence. As shown in Fig.~\ref{fig:false-color7}, each encoding step (in the downsampling part) contains two  convolutional layers with $3\times 3$ kernel size (blue arrows) with ReLU activation function and one max pooling layer with $2\times 2$ (grey arrows) kernel size, while each decoding process (in the upsampling part) contains one deconvolutional layer with $3\times 3$ kernel size (dark-green arrows), and one concatation layer to increase the dimension of the feature \cite{bell2016inside} (thin blue arrows), and two further convolutional layers. In addition, there are three consecutive fully connected layers (dash black arrows) and dropout layers (green arrows) for the depth prediction. The size of the input image is $80\times80$ pixels. The network is trained over 1000 epochs to reach an excellent convergence. This took $\sim \!\! 10$~min on a PC with a 16-core 3.50-GHz CPU, 64-GB RAM, and Nvidia GeForce RTX 3080 GPU.

\section{Results and discussion}
In following we will demonstrate the performance of the proposed method using data derived from simulations (THz-emulated MNIST \cite{Xiang2022}) and experiments, respectively.

\begin{figure*}[htb]
\centering
\includegraphics[width=\linewidth]{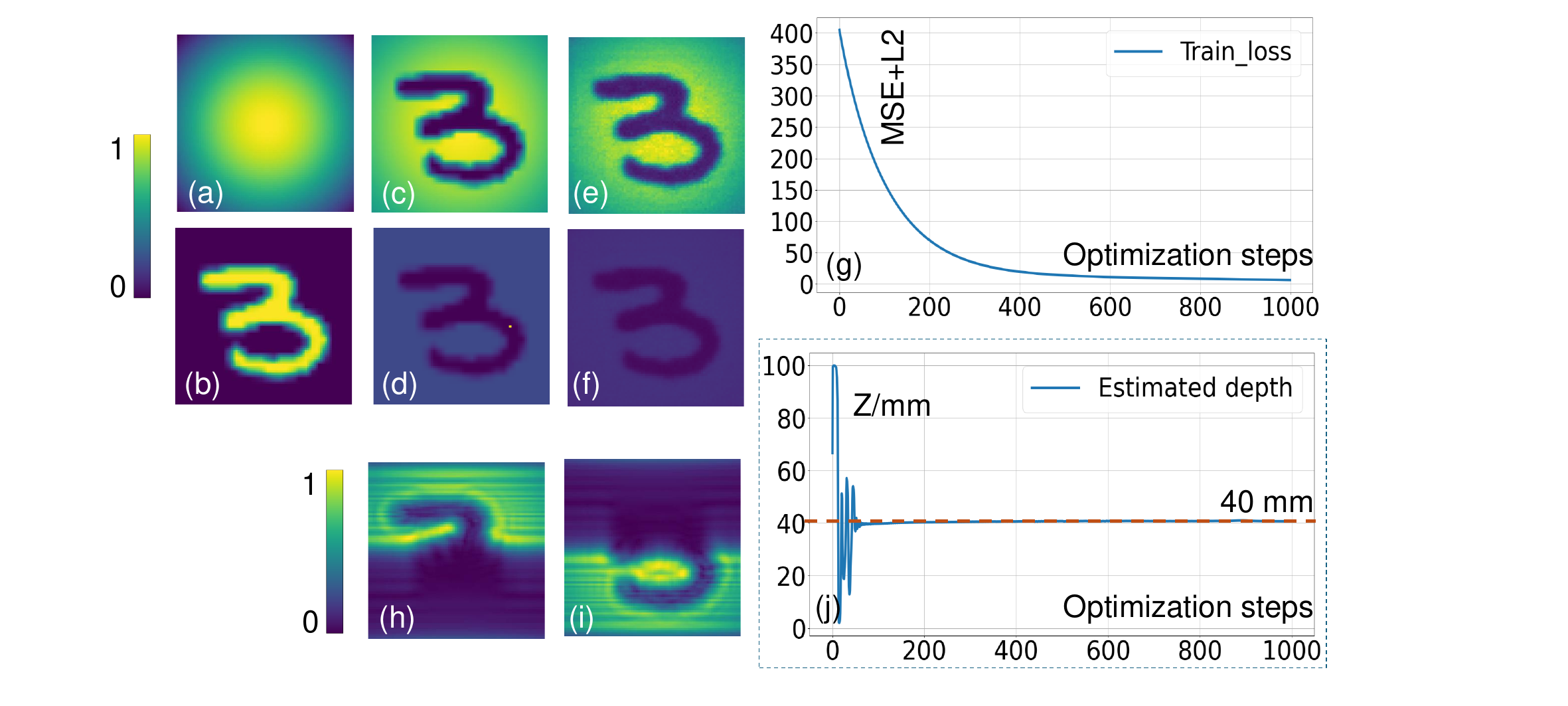}
\caption{Simulation results in the ideal situation. (a) The plot represents an ideal Gaussian intensity distribution of the THz wave; (b) one typical digit from the MNIST dataset; (c, d) ground truth of the amplitude and phase of the object; (h, i) diffraction patterns of the digit at a distance $d = 40$~mm (marked with red dashed line in the plot) with either its upper or lower half masked; (g) learning curve during the training process; (e, f) predicted amplitude and phase after training for 1000 epochs; (j) optimization curve of the depth $d$ prediction.}
\label{fig:false-color2}
\end{figure*}

In our simulation, we assumed an ideal scenario where the beam has a uniform Gaussian intensity cross-section without noise. % interference.
The frequency of the radiation was assumed to be 300~GHz.
Fig.~\ref{fig:false-color2}(a) depicts the ideal Gaussian intensity profile, with its phase being uniformly distributed. Fig.~\ref{fig:false-color2}(b) presents a digit ``3'' from the MNIST dataset, serving as an example object. The digit was assumed to be made from an infinitely thin material intransparent to THz radiation. The area around the digit was treated as having no influence on the incoming THz wave. The size of each of the $80\times80$ pixels of the object scene is 80$\times$80~mm$^2$.
The profiles of the field's amplitude and phase right behind the object, as shown in Figs.~\ref{fig:false-color2}(c) and (d), represent the ground truth of the image reconstruction process. Figs.~\ref{fig:false-color2}(h) and (i) illustrate the diffraction patterns at a distance $d = 40$~mm when occluding the upper, and the lower half of the digit, respectively. These images serve as inputs to our network. The loss curve, depicted in Fig.~\ref{fig:false-color2}(g), shows a continuous decrease along with the interactive optimization steps, stabilizing at a small value of 1.25 after 800 epochs, indicating that the network was sufficiently optimized. %At this point, the 
The resulting network's outputs include the estimated behind-the-object amplitude (Fig.~\ref{fig:false-color2}(e)) and phase profiles (Fig.~\ref{fig:false-color2}(f)). As iterations progress, the predicted depth gradually converges to a stable value of 40.01~mm %in this example 
after around 200 epochs, which is very close to the ground truth $d=40 mm$. We observed for a number of objects and distances that -- for ideal conditions -- the network's predictions for depth are all highly accurate.

\begin{figure*}[htb]
\centering
\includegraphics[width=\linewidth]{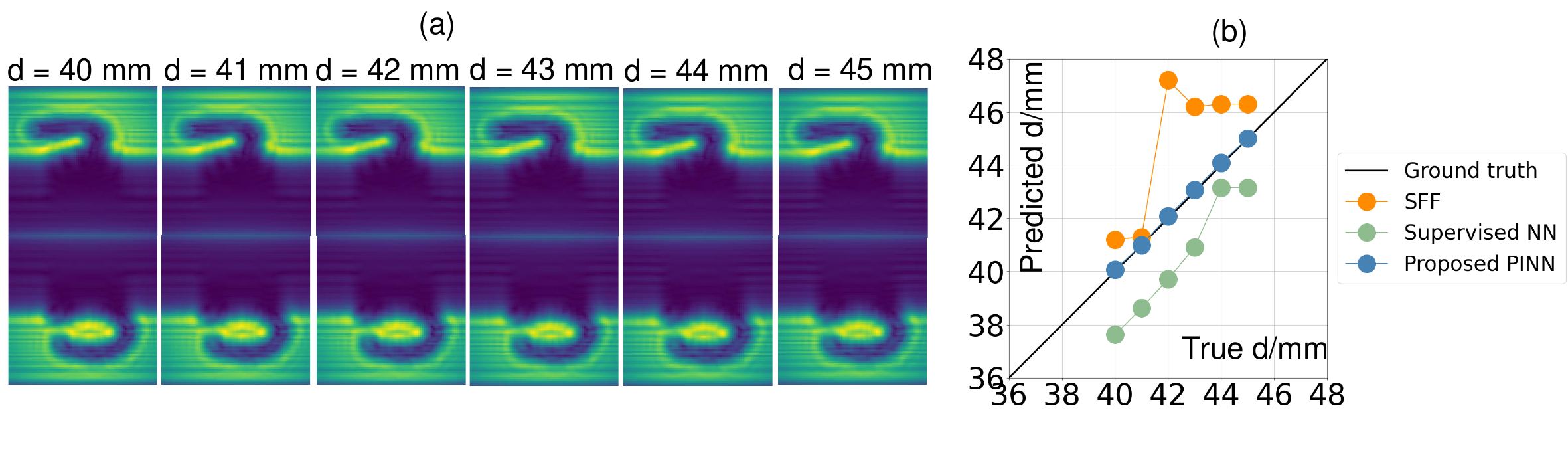}
\caption{(a) Diffraction patterns and (b) comparison of our proposed PINN with supervised CNN and SFF backpropagation for diffraction distances ranging from 40~mm to 45~mm. The line of the ground truth is nearly obscured by the data from the PINN method.}
\label{fig:false-color3}
\end{figure*}

We then compared our proposed method with (i) a combination method integrating backpropagation and shape-from-focus (SFF) calculations \cite{Perraud:19}, and (ii) supervised DL. %\HR{[You write very little about how you made the calculations. How was the SFF implemented (as you do not focus at all in your approach)? How was the DL trained? The open questions call for an additional appendix with details about the procedures.\\]}
%\MJ{I have added \nameref{AppendixB}}

Fig.~\ref{fig:false-color3}(a) shows the calculated diffraction patterns of the object with a mask over the lower (top row) and upper half (bottom row), at object distances $d$ ranging from 40 to 45~mm in steps of 1~mm.
%It is evident that at a minute diffraction distance difference of 1 mm, 
The differences in the diffraction patterns upon variation of $d$ are nearly imperceptible to the naked eye. Hence, extracting depth information from such patterns is nontrivial. %and valuable. 
Fig.~\ref{fig:false-color3}(b) compares the values of $d$ obtained by each of the three methods. % displays the comparison of our proposed PINN, supervised NN and backpropagation with SFF at diffraction distances ranging from 40 mm to 45 mm. 
The orange line represents the results of backpropagation combined with SFF (see \nameref{AppendixA} for specific sharpness curves). Its error is significant, reaching a deviation of up to %approximately 
5~mm from the ground truth. It is worth noting that for backpropagation, both phase information and amplitude are needed to generate the field, so we scale the amplitude value to the range of (-$\pi$, $\pi$) as the phase value.
%only the amplitude was utilized as input. \HR{\sout{Setting the phase to the same value as the amplitude yielded superior results compared to setting it to 1, hence here it was set to the amplitude value.} [I do not understand the message of the last sentence. How can the phase (no units) be set to the same value as the field (units: V/cm)?]} \MJ{We also need phase information to generate the filed, so the amplitude value is scaled to the range of (-$\pi$, $\pi$) as the phase value.}
The green line represents the predictions of the supervised NN, which performed slightly better than backpropagation combined with SFF, the former, however, consistently predicting lower values of $d$ than the ground truth, and the latter, in contrast, higher values. At $d=44$~mm and $d=45$~mm, %\HY{[This should be $d=44$~mm and $d=45$~mm from the figure? yes I have revised it]} 
the supervised NN determined identical values of $d$. This is because in preparing the training dataset, it was impractical to cover all possible depth positions, because the depth values are generated using a random generator (\nameref{AppendixB}),%\HR{[which values of $d$ were covered (info to be given in the appendix?)?]}, \MJ{because the depth value are generated using random generator (\nameref{AppendixB}), }
resulting in potential information gaps at certain depths. Our supervised network was trained with 60,000 pairs of digits and 10,000 pairs were used as testing datasets. Theoretically, its predictive performance should improve with further data augmentation.
The blue line in Fig.~\ref{fig:false-color3}(b) presents the predictions of our proposed hybrid multi-head PINN method, which agree very well with the ground truth data (black line). The proposed method significantly outperformed the other two. Moreover, this method doesn't require time-consuming pre-training like supervised learning.

We further evaluate the coefficient of determination ($R^2$) to assess the performance of depth prediction from the three methods. The $R^2$ value can be calculated as follows \cite{miles2005r}:
\begin{equation}
R^2 = 1 - \frac{SS_{res}}{SS_{tot}}\,,
\label{eq:refname6}
\end{equation}
where $SS_{res}$ is the sum of the squares of each ground truth and model prediction, and $SS_{tot}$ is the sum of squares of each ground truth and the mean of the ground truths. %\HY{squares of the differential value of the predictions and ground truths?}
%\HR{[squares of what quantity?]}
Generally, the closer the $R^2$ is to 1, the better the predictive performance of the model. The method proposed in our work outperforms the other two methods obviously: the $R^2$ values for backpropagation with SFF, supervised learning, and our PINN method are -1.6168, -0.4291, and 0.9986, respectively.

\begin{figure*}[htb]
\centering
\includegraphics[width=\linewidth]{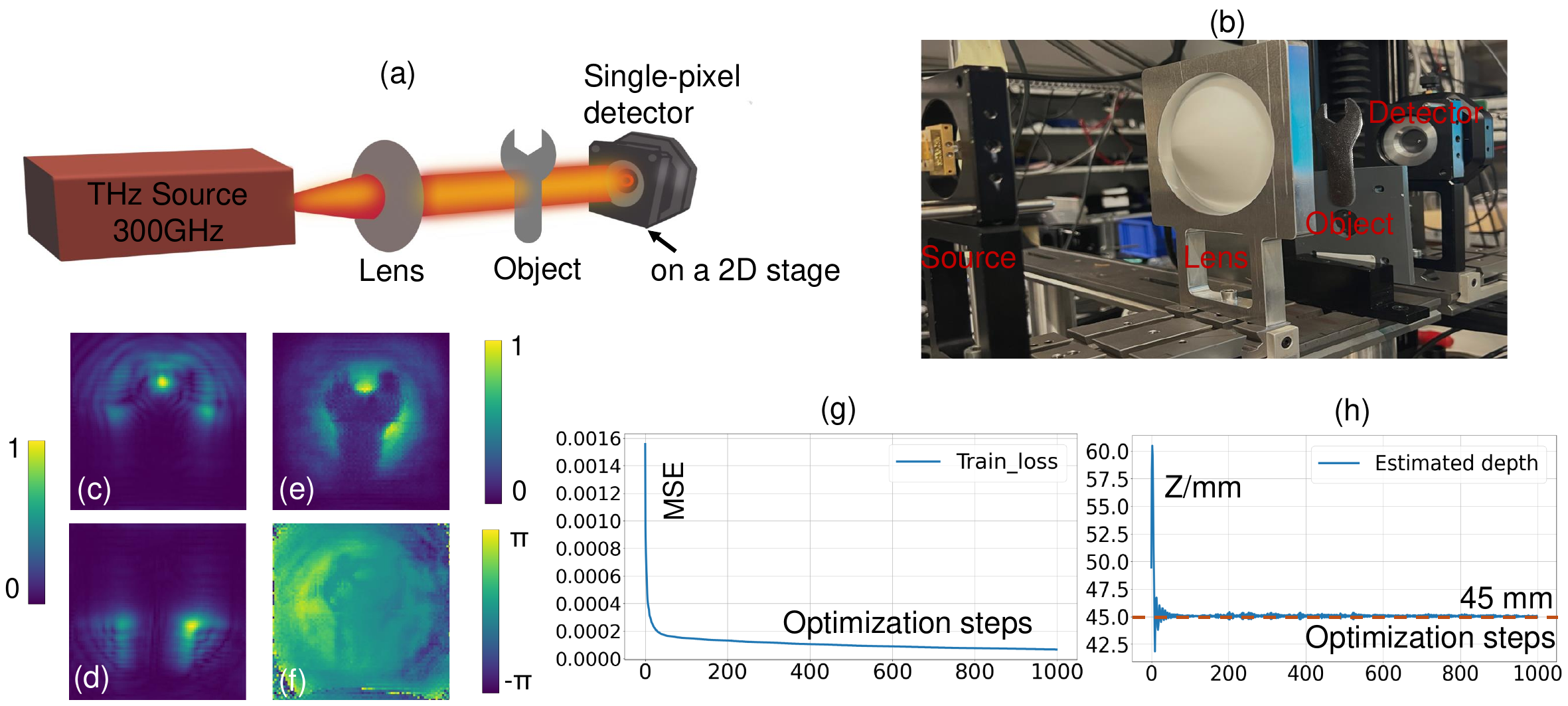}
\caption{Experimental setup and results. % in realistic situation. 
(a, b) Schematic and photograph of the THz in-line-holographic imaging system (LO-wave injection not shown), (c, d) measured diffraction patterns of a wrench, (e, f) reconstructed amplitude and phase, (g) loss curve during the training process, (h) optimization curve of the depth estimation, where the ground truth is $d=45$~mm (marked with red dashed line in the plot).}
\label{fig:false-color4}
\end{figure*}

To validate the proposed method in realistic scenarios we employed experiment-derived data, which consists of measured diffraction patterns of objects at 300~GHz using a setup for in-line holography (single-beam lens-less holographic imaging \cite{Wan2019, Xiang2022}) with heterodyne detection \cite{Glaab2010, Boppel2012}. %\HR{\sout{In a first experiment, we acquired the diffraction pattern by power detection using the setup displayed in Fig.~\ref{fig:false-color4}. Later, we switched to heterodyne detection  because images obtained in that way are much less affected by noise than those recorded by power measurements.}} 
A schematic diagram and a photograph of the measurement setup are shown in Figs.~\ref{fig:false-color4}(a) and (b), respectively. The set-up is described in detail in \nameref{AppendixC}. 

As an object, we used a metallic wrench placed at a distance $d=45$~mm from the object. The wrench had a thickness of 1~mm and a lateral size of around $30\times60$~mm$^2$; the scanned image area was $80\times80$ pixels with a 1-mm$^2$ pixel pitch. 
Two recorded diffraction patterns of the object are shown in Figs.~\ref{fig:false-color4}(c) and (d), with the lower, respectively upper half of the object being masked. 

As an input for our PINN, we used the amplitude maps. The phase maps were utilized independently to determine the precise distance of the object from the detector (given the housing of the detector and the substrate lens on it, a measurement with a caliper is not precise enough). The correct distance was derived by numerical backpropagation of the complex-valued diffraction patterns, evaluation of the sharpness \cite{perraud2018IRMMWTHz, nayar1994shape, Yuan2023} of the images at different distance values chosen during the reconstruction, and final selection of the distance for which the largest sharpness is obtained. 
%[Is this correctly described?] }
%\MJ{correct!}

% In realistic scenarios, unlike the ideal case where a uniform Gaussian beam is assumed as discussed earlier, there may be non-uniform distributed noise present. 

If the algorithm represented schematically in Fig.~\ref{fig:false-color1} was used with real data, the reconstructed images could be of fairly poor quality and the depth estimation to be off from the real value. It turned out that this was mainly due to deviations of the cross-sectional amplitude and phase profiles of the real THz beam from the ideal case. The algorithm, unable to distinguish between the beam and the object, imposed beam imperfections onto the reconstructed object.

This problem was remedied in the following way. Using heterodyne imaging, we measured amplitude and phase of the THz beam (in the following called the ``reference'') without object, see \nameref{AppendixC}. The amplitude and phase of the reference beam then served as prior information. It was used in the physical model $H$, which was modified in that way that the outputs of the amplitude and phase heads of the PINN were multiplied with the measured amplitude, respectively phase profile, before performing the propagation calculations of the numerical diffraction simulation with the original model $H$.\footnote{More precisely, we used  the sigmoid activation function to ensure the amplitude and phase outputs to be between 0 and 1, then multiplied them by the reference to get the final amplitude and phase predictions.} 
This was repeated in each epoch. While leading to a large MSE error after the first epoch, this approach led the PINN in later iterations to process only the object data, free from the specifics of the object-illuminating beam. Concomitantly, the depth estimation improved.

Fig.~\ref{fig:false-color4}(g) illustrates the loss curve of our reconstruction, showing that after 500 iterations the network's loss stabilized at a value of 0.000085. It is worth noting that, unlike the theoretical case discussed above, we calculated the loss in Eq.~\ref{eq:refname5} without L2 regularization.
As stated before, the regularization is important to prevent overfitting if a scene is of sufficiently low complexity. With the experimental data, we found however, that the imperfections of the illumination itself create sufficient complexity such that the regularization is not needed anymore. When keeping the regularization, the convergence was slower and the results not better than without the regularization.
% because \HR{\sout{the noise is non-Gaussian} the reference deviated from a perfect Gaussian. [Is this the correct statement? Do you mean ``noise´´ or deviations of the reference beam (and its phase profile) from a Gaussian? Please note that I avoided the word ``noise'' because noise has a time dependence, while the beam imperfections discussed above do not. By the way, for my curiosity: Does the discarding of the L2 regularization make a big impact or is it a small matter?]} \HR{\sout{This also speeded up the convergence of the network,} [It is not clear what helped to speed it up: the use of the reference beam or the neglect of the L" regularization. And compared to what did it speed up? Compared to the ideal case or to the case without taking into account the reference?]} \HR{\sout{indicating that there is only a 0.000085 MSE between the actual diffraction patterns (input) and the diffraction patterns predicted by the integration of the network and physical model $H$'s evaluation.}} 

Figs.~\ref{fig:false-color4}(e) and (f) depict the network predicted amplitude and phase maps, respectively. Fig.~\ref{fig:false-color4}(h) shows the depth estimation from the network as a function of the optimization steps (iterations). Clearly, the depth prediction from the network quickly stabilizes onto the correct depth value ($d = 45$~mm), slightly varying around that ground truth value with the iterations. 

As a comparison, we also evaluated the depth prediction performance with the method combining backpropagation and SFF, as well as the prediction performance from supervised learning method. However, both of these two approaches failed to accurately predict the distance.

\begin{figure*} [htb]
\centering
\includegraphics[width=\linewidth]{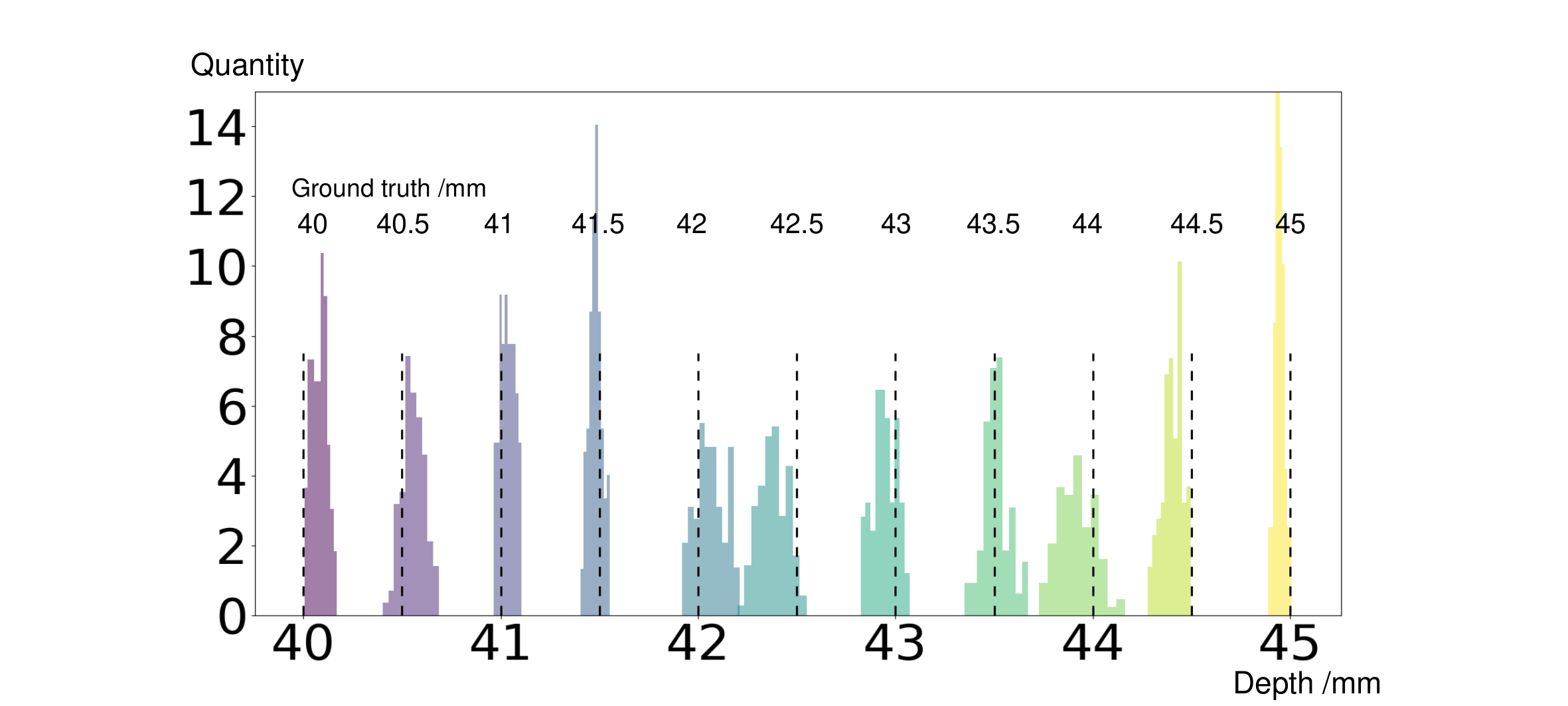}
\caption{Distribution of the predicted depth, representing distance resolution of our proposed PINN at diffraction distances ranging from 40~mm to 45~mm (500 predictions of a given fixed depth were used to generate each distribution in the figure) with ground truths marked with dashed lines.}
\label{fig:false-color5}
\end{figure*}

Given that our method, in contrast, can predict the depth information well %accurately predict depth information 
with minimal errors in both the %ideal and realistic 
theoretical and experimental scenarios, we proceeded to determine its distance resolution for the ideal case. According to the theory of object reconstruction from diffracted amplitude and phase data of an object at an unknown distance $d$ from the detector, %.of phase detection, 
the depth resolution can be quantified as \cite{Yuan2023}:
\begin{equation}
%\HY{\sout{\sigma_d = 4\gamma \frac{\lambda d^3}{D^3}}} 
\sigma_d = 4\gamma \frac{\lambda d^2}{D^2}.
\label{eq:refname8}
\end{equation}
%\HY{[Here I changed it to the ratio of the square of $d$ and $D$. In Ref. \cite{Yuan2023} page84, the lateral resolution should be defined by limited area case as Eq.(5.13), $\sigma=\gamma\frac{\lambda d}{D_{max}}$, there $D_{max}$ should be the 1D size of the recording area, in this paper we define as $D$. Later on Page91, the depth resolution should be substituted by using $\sigma_d = \frac{4d}{D}\sigma$ in the text of the last paragraph of Sec. 5.2.1. There $D$ (defined as aperture) should also be the recording area $D$ in this paper. Therefore the resultant derivation should be above the formula. I suggest using the minimum 1D value to calculate (80 mm) to give the resolution. Mingjun, could you go back to the Ref to double check if I defined it correctly?]}
Here, $D$ represents the one dimensional minimum size of the recording area (in our case with the chosen dataset, it is 80~mm) and $\lambda$ denotes the wavelength of the radiation. When the phase difference between pixels on objects is zero, $\gamma = 1.64$ \cite{Yuan2023}. Therefore, in our system, the theoretical resolution can be determined to be around 1.64~mm. 
%\HR{[Why do you give a range of values and not one value? I obtain 1.17~mm with the numbers which you specify. You cannot give a range without explanation.]} 

%\MJ{So here the resolution is 1.64~mm}

%\HR{[For the following paragraph: Mingjun, you didn't specify whether the statistics was obtained for the ideal case or the experimental one. I wrote assuming that the former is the case.}
%\MJ{I obtained the data at 45mm and then back-propagated to z=0, and then propagate the data to at z=40,40.5....44.5mm}

In order to compare this prediction\cite{Yuan2023} with the depth resolution obtained statistically from repeated `measurements' using our trained PINN, we calculated the diffraction patterns for distances from 40~mm to 45~mm, with interval of 0.5~mm, in total 11 distances. 
%We collected an amplitude diffraction pattern at $z=45mm$, then back-propagated to $z=0$, and subsequently forward-propagated to generate a dataset at distances from 40 to 45~mm at 0.5~mm intervals. 
We used these patterns as test inputs to the trained PINN for a resolution evaluation. To demonstrate the robustness, we conducted 500 predictions for each distance. 
Fig.~\ref{fig:false-color5} displays the histogram of the predicted depth results for each distance. %, where the horizontal axis represents the depths and 
The vertical axis represents the histogram of counts for PINN predictions over each single depth. It can be observed that the PINN can accurately predict diffraction patterns with roughly 0.5~mm error at most, and its resolution can even be better than 0.5~mm as shown in the picture. %\HR{[The 42.5-mm predictions have an interesting offset to lower distance! Any idea why this data set behaves a bit different than the others?]}
% \HR{[HUI: YOU HAVE TO LOOK INTO THIS: How can a CNN beat the physical diffraction limit. Please add text (a comment). From a fundamental point of you, the DL approach also relies on the phase for the depth information, so there should apply related rules. I can only imagine that 
The resolution -- in this noise-less and aberration-less theoretical scenario -- hence is better than that predicted resolution from Eq.~\ref{eq:refname8}. The proposed hybrid multi-head PINN method appears to be more sensitive to changes in the interference patterns induced by different depth, thus delivering better resolution in determining the depth than those predicted by the Rayleigh criterion which had been used in deriving Eq.~\ref{eq:refname8}. This is a very promising finding for further future experimental exploration. 

\section{Conclusions}
In conclusion, we have developed a novel physics-informed deep learning method for predicting the depth of planar objects using just two diffraction patterns obtained by masking the upper and lower halves of the object. This was motivated by the failure using just a single diffraction pattern for the object without masking (see more details for the case without masking in \nameref{AppendixD}).
%\HR{[Now thinking about it: It would have been good if we would have made the case why the masking is needed. We do not prove the need. Suggestion: Can we show the loss curve for a case without masking, say, in an appendix?]} \MJ{(see more details for a case without masking in \nameref{AppendixD})}
In contrast to supervised learning methods that require large training datasets to optimize their weights and biases, and optical approaches involving backpropagation and SFF, our method offers more accurate and stable predictions in both ideal and realistic scenarios. %Furthermore, we have investigated the resolution of depth reconstruction and found that the resolution of this 
Additionally, a numerical investigation into the attainable resolution of distance reconstruction demonstrates that our method can achieve a theoretical resolution of 0.5~mm or even better, surpassing the resolution as determined by the Rayleigh criterion. This reveals the high sensitivity of the neural network to small changes of interference patterns induced by even tiny changes of the object distance. Therefore, our hybrid multi-head physics-informed neural network significantly advances the depth estimation for planar objects in THz holographic systems, marking
%can achieve the estimation of depth information for planar objects in the THz holographic system, signifying 
a crucial step towards THz 3D imaging.

\section*{Acknowledgment} 
The work is supported by (i)XFI-JRC collaboration fund (M. Xiang); (ii) the German Research Foundation DFG with grant RO 770/48-1 (H. Yuan); (iii) the BMBF funded KISS consortium (05D23RI1) in the ErUM-Data action plan (K. Zhou), (iv) the CUHK-Shenzhen university development fund under grant No. UDF01003041 and UDF03003041, and Shenzhen Peacock fund under No. 2023TC0179 (K. Zhou).

% \section*{Appendix A}

% \begin{figure*}[htb]
% \centering
% \includegraphics[width=\linewidth]{nn}
% \caption{Structure of the proposed hybrid multi-head PINN}
% \label{fig:false-color7}
% \end{figure*}
% The architecture of the proposed PINN as employed for the supervised DL method is implemented using TensorFlow, an open-source deep-learning software package \cite{abadi2016tensorflow}. We use TensorFlow version 1.9.0 and Python 3.6.5. We adopt the estimation (Adam) \cite{kingma2014adam} with a learning rate of 0.0005 to optimize the weights and biases. We add uniformly distributed noise between 0 and $\frac{1}{30}$ to the fixed input diffraction pattern in every optimization step to achieve better convergence. As shown in Fig.~\ref{fig:false-color7}, each encoding step contains two 3$\times$3 convolutional layers (blue arrows) with Relu activation function and one 2$\times$2 max pooling layer (grey arrows), while each decoding process contains one 3$\times$3 deconvolutional layer (dark-green arrows), one concat layer to increase the dimension of the feature \cite{bell2016inside} (thin blue arrows), and two convolutional layers. In addition, there are three consecutive fully connected layers (dash black arrows) and dropout layers (green arrows) for depth prediction. The size of the input image is 80$\times$80 pixels. The network usually needs 1000 epochs to find an excellent estimation. This takes ~10 min on a PC with a sixteen-core 3.50-GHz CPU and a 64-GB RAM, using Nvidia GeForce RTX 3080 GPU. 

\section*{Appendix A}
\label{AppendixA}
\begin{figure*}[htb]
\centering
\includegraphics[width=\linewidth]{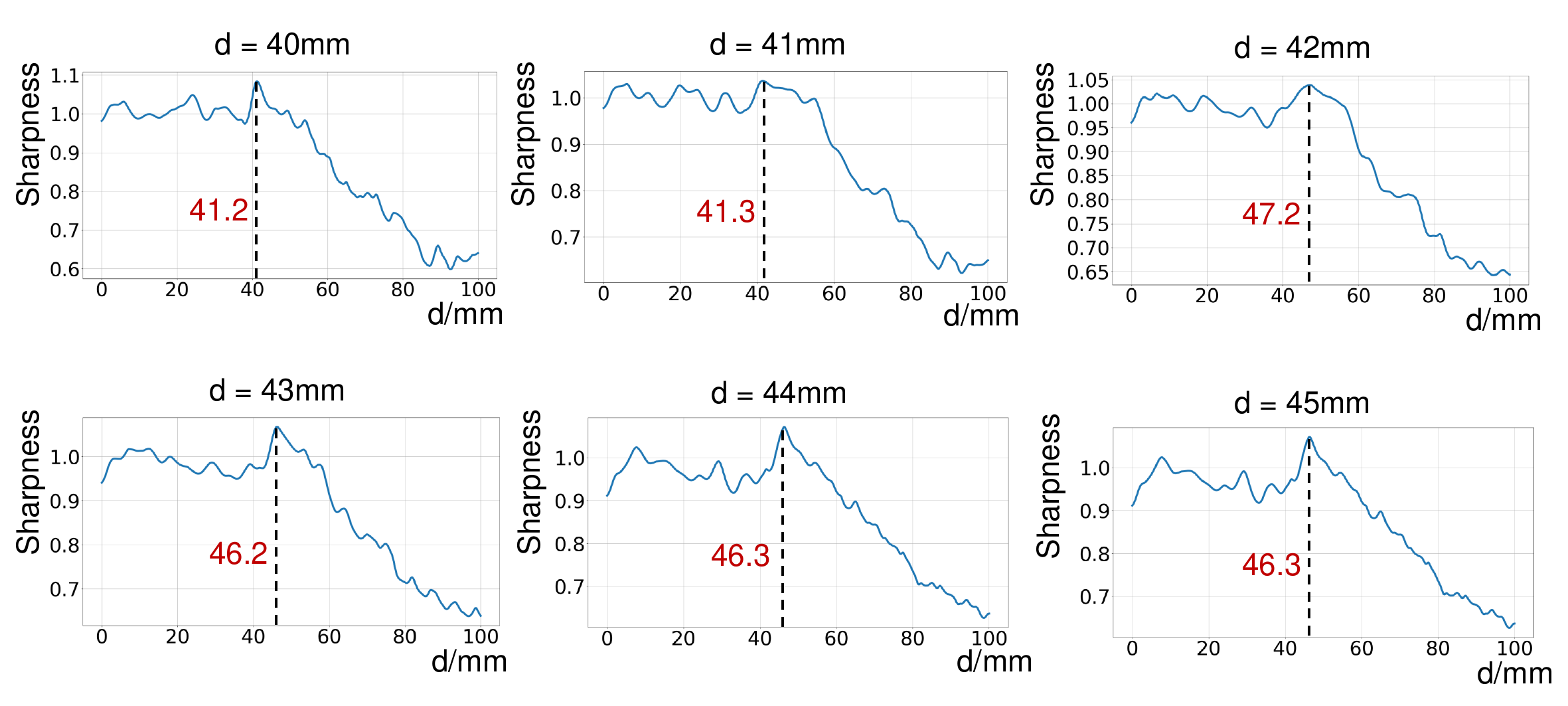}
\caption{SFF results at various diffraction distances from 40~mm to 45~mm.}
\label{fig:false-color8}
\end{figure*}
Figure~\ref{fig:false-color8} presents sharpness curves derived using the SFF algorithm. The SFF algorithm uses the sum of the modified Laplacian (ML) operator described in \cite{nayar1994shape} on each pixel of all the images to calculate the position of the focus point on the optical axis. 
%\HR{[What is ``focus versus the optical axis''? Do you mean ``position of the focus point on the optical axis''?]} 
Then, a focus measure at the pixel $(x,y,z)$ is calculated by: 
\begin{equation}
F(x,y,z) = \sum_{X=x-N}^{x+N} \sum_{Y=y-N}^{y+N} ML(X,Y,z) \quad if \, ML>T
\label{eq:refname9}
\end{equation}
$N$ determines the window size to compute the focus measure. Here we set $N=1$. $T$ is a threshold value to remove the contribution of regions of the image with too low contrast. The depth position is found by the  maximum value $d = z|_{max(F(x,y,z))}$ where the contrast is sharpest.

\section*{Appendix B}
\label{AppendixB}

\begin{figure*}[htb]
\centering
\includegraphics[width=\linewidth]{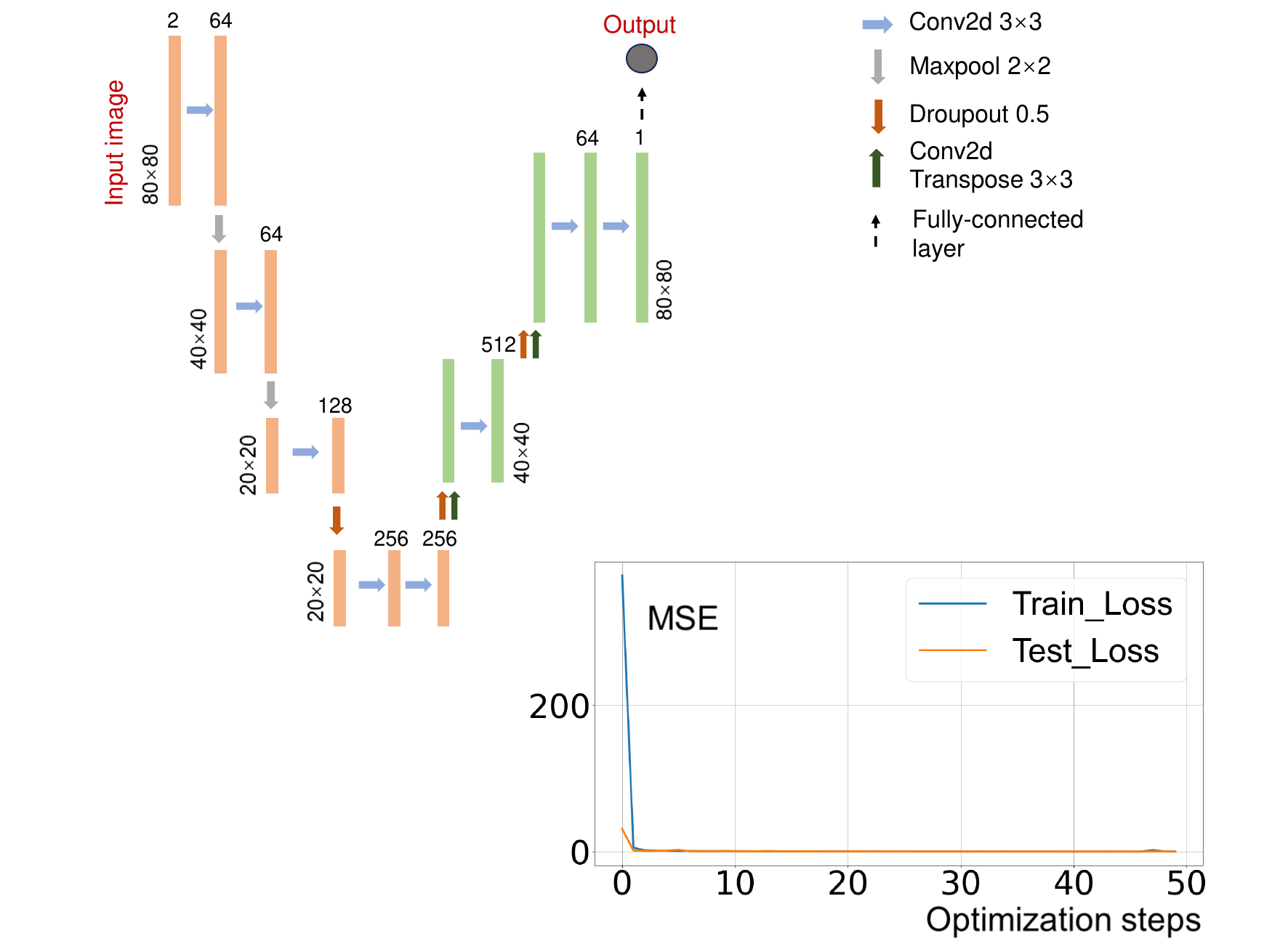}
\caption{Structure and training curve of the supervised NN}
\label{fig:false-color10}
\end{figure*}
Figure~\ref{fig:false-color10} shows my supervised NN. I used the MNIST dataset to generate training and testing data, consisting of 60,000 training data and 10,000 testing data. The specific process is as follows: First, each original 28x28 pixel digit image is enlarged to 80x80 pixels. Then, the enlarged image is multiplied by the amplitude of the Gaussian beam (Fig~\ref{fig:false-color2}(a)) to obtain amplitude, and by the phase of the Gaussian beam to obtain phase information.
Next, to simulate the effect of being partially obscured by metal, I set the upper or lower half of each image to zero. After that, diffraction patterns of the obscured upper and lower halves are calculated using the diffraction model H. It is important to note that each image uses a random number between 0 and 100~mm as the depth information.
Finally, the two diffraction patterns obtained are combined into a matrix of shape (80, 80, 2) and used for network training along with the depth information. This combined matrix contains information on both parts, allowing the network to learn the diffraction characteristics of both amplitudes and thus achieve depth matching.
It can be observed that as the number of optimization iterations increases, both the training loss and the test set loss (with the loss function defined in Eq.~\ref{eq:refname4}) decrease, indicating that the network is improving without overfitting. We then use this trained network to predict the diffraction spectrum. The results are shown by the green line in Figure~\ref{fig:false-color3}(b).

\section*{Appendix C}
\label{AppendixC}
\begin{figure*}[htb]
\centering
\includegraphics[width=\linewidth]{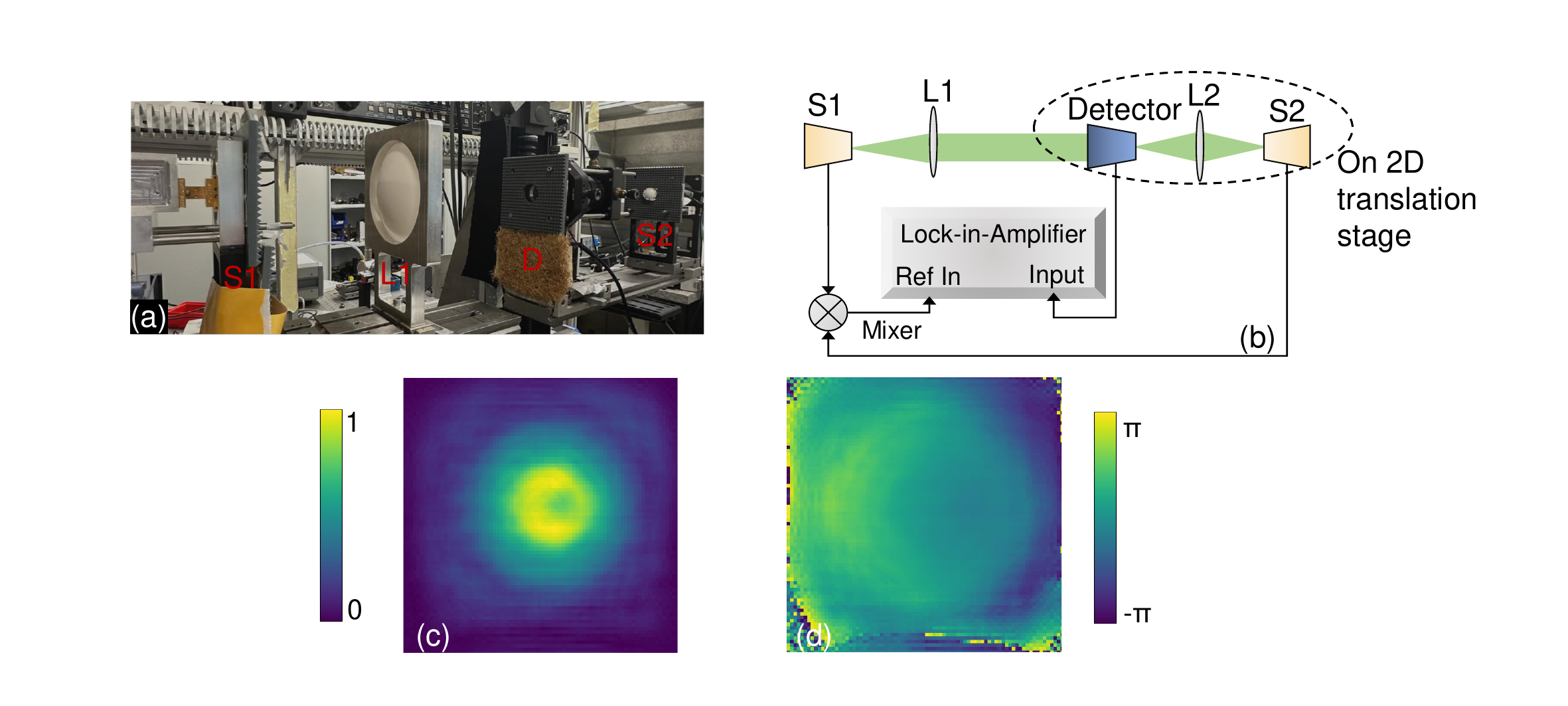}
\caption{Experimental heterodyne setup and detected reference images. (a, b) Photograph and schematic representation of the THz in-line holographic imaging system with heterodyne detection, (c, d) cross-sectional amplitude and phase images of the illuminating beam recorded by the heterodyne detection system.}
\label{fig:false-color9}
\end{figure*}

A photograph and a schematic of the experimental heterodyne setup are shown in Figs.~\ref{fig:false-color9}(a) and (b). We used this setup to obtain the ground truths. The system \cite{Yuan19, 9370667} contained two frequency-locked 300-GHz electrical multiplier chains (S1, S2) with a 18-kHz frequency offset between them. One (S2, vendor: RPG-Radiometer Physics GmbH; output power: 432~$\mu$W) was used as a local oscillator (LO) and the other (S1, vendor: Virginia Diodes, Inc.; output power: 1~mW) provided the object-illuminating wave. The LO wave was focused onto the detector directly by a biconvex focusing lens, L2, with a 5-cm focal length and 2-inch aperture. The illuminating radiation was collimated by a focusing lens, L1, with a 10-cm focal length and 4-inch aperture. %L1 was an aspheric lens. 
Given the required large NA, its aspheric hyperboloidal-planar shape avoided the spherical aberrations that one would encounter if a spherical lens had been used. The aperture of L1 had to be large enough to guarantee 
that the objects were illuminated over their full extension with a collimated Gaussian beam free of beam amplitude modulations arising by diffraction at the edge of the lens. Such modulations introduce high-spatial-frequency signatures which can make the learning process difficult. 
Furthermore, the alignment of the lens was also crucial, since the phase measurement is sensitive to optical-axis displacements. Both L1 and L2 were homemade from Teflon.

The detector was a single-pixel TeraFET detector \cite{Ikamas2018}, mounted on a 2D translation stage to scan and record the diffraction patterns. The radiation arriving from the object entered the detector chip from the back-side, passing a Si substrate lens with a diameter of 4~mm \cite{PIERS19, IWMTS22}. The LO radiation was coupled in from the front-side through a superstrate lens \cite{Yuan19, yuan2023-lith}.

Fig.~\ref{fig:false-color9} (c) and (d) show amplitude and phase maps of the beam, demonstrating that the beam had an isotropic amplitude distribution and near-constant phase over its cross-section. A numerical evaluation of the beam profile revealed a Gaussian-like beam shape. The bluish horizontal segments in Fig.~\ref{fig:false-color9}(d) result from position shift errors induced by forward- and backward-scanning of the translation stage. This effect is not visible in the amplitude map because the amplitude is insensitive to slight variations of the scanner position. This also proves that our algorithm has no strict requirements for the calibration of the reference beam and can be used in a simple way.

\section*{Appendix D}
\label{AppendixD}

\begin{figure*}[htb]
\centering
\includegraphics[width=\linewidth]{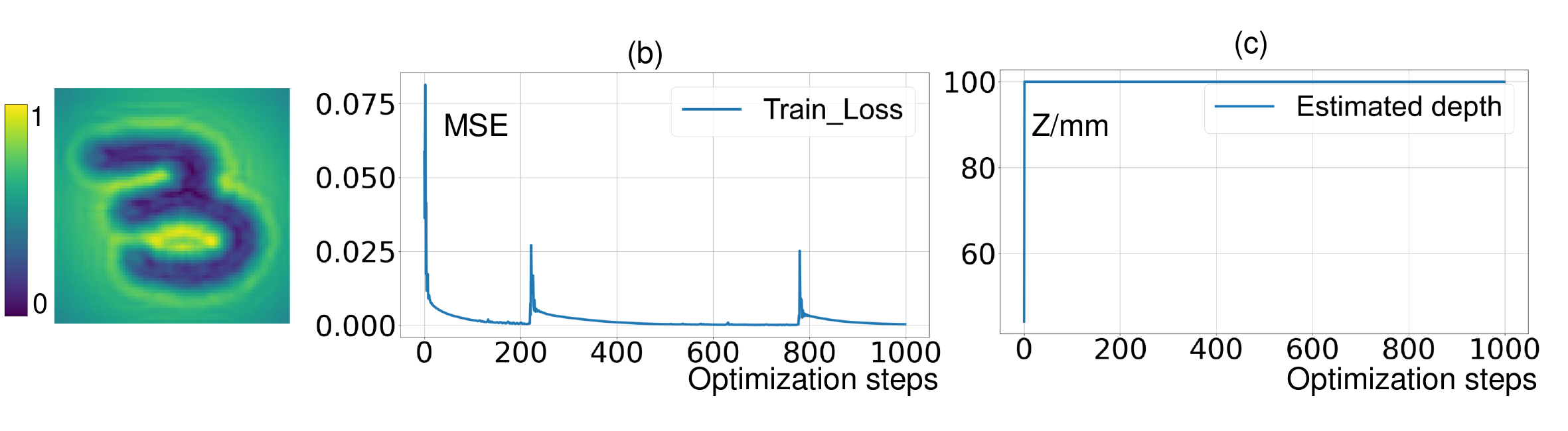}
\caption{Depth prediction using one diffraction pattern}
\label{fig:false-color11}
\end{figure*}
Initially, we aimed to use a single diffraction pattern to predict depth information. Although the loss function gradually decreases, the predicted depth converges to a local optimum and stops improving then, resulting in inaccurate outcomes. As shown in Figure~\ref{fig:false-color11}, the left panel is the single diffraction pattern taking as input to the network, (b) illustrates the loss curve during training, which overall shows a downward trend, and (c) displays the predicted depth values, indicating that the depth quickly converged to a wrong number 100~mm and no longer get adjusted, while the ground truth is 45mm. Subsequently, we tried occluding the upper or lower part of the object and used two images as input, which significantly improved the results, as shown in Figure~\ref{fig:false-color2} in the main text.

\bibliographystyle{elsarticle-num}
\bibliography{cip-v3-bst-references}
\end{document}